\documentclass[a4paper,10pt]{article}
\usepackage{amsmath,amssymb,epsfig,graphicx} 
\usepackage{setspace}

\begin{document}


\title{ME-PS matching in the simulation of multi-jet production 
in hadron collisions using a subtraction method}

\author{Shigeru Odaka, Norihisa Watanabe, and Yoshimasa Kurihara\\
High Energy Accelerator Research Organization (KEK)\\
1-1 Oho, Tsukuba, Ibaraki 305-0801, Japan\\
E-mail: \texttt{shigeru.odaka@kek.jp}}

\date{}

\maketitle

\begin{abstract}
The subtraction method for the matching between the matrix element (ME) 
and parton shower (PS), which has been developed for combining 0-jet and 
1-jet production processes in association with electroweak-boson production 
in hadron collisions, is extended to multi-jet production.
In order to include multi-jet MEs, we have to address 
the soft-gluon divergence together with the collinear divergence.
We introduce an approximation that simultaneously reproduces both 
divergences in a form suitable for application to our subtraction method.
The alteration in the subtraction can be compensated by applying 
an appropriate correction to corresponding non-radiative events.
We demonstrate that $W$ + 0, 1, and 2 jet production processes can be 
consistently combined using the developed matching method.
\end{abstract}

\section{Introduction}
\label{sec:intro}

As a result of higher-order quantum chromodynamics (QCD) interactions, 
hard interactions producing large transverse momentum particles 
or heavy particles, {\it e.g.}, weak bosons, are frequently associated with 
hadron jets in high-energy hadron collisions such as the proton-proton 
collisions provided by CERN LHC.
A precise understanding of the jet production is necessary not only to 
achieve precise measurements of hard interactions but also to reliably 
estimate the background of unknown new phenomena.
Since the jet production and other soft QCD interactions complicate 
the produced events, 
simulations based on Monte Carlo (MC) event generators are 
indispensable tools for measurements at hadron colliders.
Hence, it is important that the jet production is precisely reproduced 
in MC event generators.

The hadron jets are considered to originate from the production of 
energetic partons (light quarks and gluons) 
that can be evaluated in the framework of the perturbative QCD (pQCD).
Multi-jet production can be simulated by parton showers (PSs) 
in MC event generators.
However, since the PS is based on the collinear approximation of pQCD, 
we cannot expect satisfactory precision for the production of isolated 
energetic jets that mimic the signature of various new phenomena.
Simulations based on the matrix element (ME) including the jet 
(energetic parton) production are necessary for reproducing such phenomena.

We have to apply PS simulations also to events generated according to 
jet-including MEs in order to provide realistic simulations.
A problem arises when we simultaneously employ ME and PS for the jet production.
There may be an overlap (double-count) between the two simulations.
The problem is serious if we try to construct a consistent simulation 
by combining simulations based on MEs of different jet multiplicities.
PS simulations applied to lower-multiplicity events may overlap with 
the jet production in higher-multiplicity MEs.
The double-count can be avoided by separating the phase space to be covered 
by the PS and ME.
However, such a separation may produce discontinuities in observable 
spectra because the ME includes non-divergent components 
that are ignored in the PS.

Several solutions to this problem have been proposed and implemented 
in event generators:
the ME correction in PYTHIA~\cite{Miu:1998ju} and HERWIG~\cite{Seymour:1994df},
the CKKW method~\cite{Catani:2001cc} implemented in 
SHERPA~\cite{Gleisberg:2008ta}, 
and a subtraction method in MC@NLO~\cite{Frixione:2002ik}.
The MLM prescription in AlpGen~\cite{Mangano:2002ea} can be considered 
as a variation of the suppression method of CKKW, 
and POWHEG~\cite{Nason:2004rx} is employing another suppression method.
Among them, multi-jet ($\geq 2$ jets) MEs can be included by the CKKW method 
and the MLM prescription only, 
and both are based on the suppression method
in which multi-jet events generated according to jet-including MEs are 
suppressed by reinterpreting them in the picture of a PS.
This technique has been refined and implemented in various MC event 
generators~\cite{Alwall:2008qv,Hoeche:2009rj,Lonnblad:2012ng}.

We have previously developed a technique to achieve the matching between 
ME and PS (ME-PS matching) for the jet production 
on the basis of a subtraction method~\cite{Kurihara:2002ne,Odaka:2007gu}, in 
the framework of the GR@PPA event generator~\cite{Tsuno:2002ce,Tsuno:2006cu}.
We have established a matching method for combining the simulations 
based on MEs including 0-jet and 1-jet production in association with 
the production of weak boson(s): $W$, $Z$, $W^{+}W^{-}$, $WZ$, and 
$ZZ$~\cite{Odaka:2011hc,Odaka:2012da}.
The method has also been successfully applied to diphoton ($\gamma\gamma$) 
production~\cite{Odaka:2012ry}.

The source of the double-count is the divergent leading-logarithmic (LL) 
component in 1-jet MEs.
This component is identical to the leading term of the PS that is applied 
in the simulation based on 0-jet MEs.
In our matching method, we numerically subtract the LL components from 
1-jet MEs.
Since the divergent components are subtracted, 
the remaining cross sections are all finite without introducing any cutoff.
The PS simulation is usually limited by a certain energy scale.
Accordingly, the subtraction is limited at the same energy scale.
Hence, we call this method the limited leading-logarithmic (LLL) subtraction.

Our event generators are composed of tree-level MEs and primitive 
virtuality-ordered LL PSs.
Although the merging of next-to-leading order 
MEs~\cite{Hamilton:2012rf,Alioli:2012fc}\footnote{See also the references therein.} 
and the merging of multi-jet MEs for developing next-to-next-to-leading 
order event generators~\cite{Hamilton:2013fea,Karlberg:2014qua} 
are recently discussed, 
the primary purpose of MC event generators is to provide tools 
for experimental measurements.
For such use, precision in kinematical distributions of generated events 
is the primary demand 
while precision in the absolute value of cross sections is less important.
We have shown in previous studied~\cite{Odaka:2009qf,Odaka:2012iz,Odaka:2013fb} 
that simulations with our event generator reproduce the production kinematics 
of $Z$ bosons very precisely, 
and that the application of an appropriate PS-branch kinematics model 
is crucial to achieve the precision.
It must be worthwhile to advance the development to more complicated 
multi-jet processes, 
at least to clarify the capability of the simulations with our simple setup.
Such studies will help us to understand the actual benefits of 
higher-order corrections.

In this article, we extend the LLL subtraction method to 2-jet 
production processes.
In order to deal with multi-jet ($\geq 2$ jets) production processes, 
we have to account for the soft-gluon divergence together with the 
collinear divergence approximated by the LL component in the LLL subtraction.
We introduce an approximation that simultaneously reproduces the 
two divergences.
We also introduce a correction to PS-applied 1-jet events according 
to this alteration in the subtraction.
We subsequently demonstrate that the developed matching method provides 
a reasonable simulation of $W$ + jet production up to two jets.

The rest of this article is organized as follows.
We introduce a subtraction method that simultaneously subtracts the collinear 
divergence and the soft-gluon divergence in Sec.~\ref{sec:combsub}.
The correction to PS-applied events that compensates for the alteration 
in the subtraction is described in Sec.~\ref{sec:pscorr}.
The performance of the ME-PS matching method employing the developed techniques 
is tested for $W$ production processes in Sec.~\ref{sec:matchedgen}, 
and the discussions are concluded in Sec.~\ref{sec:concl}.

\section{Combined subtraction}
\label{sec:combsub}

The collinear approximation that is subtracted from the squared MEs of radiative 
processes in the LLL subtraction can be expressed as~\cite{Odaka:2011hc} 
\begin{equation}\label{eq:collapp}
  T^{\rm coll} = \sum_{a} T^{\rm coll}_{a} , \ \ \
  T^{\rm coll}_{a} = { 8\pi\alpha_{s} \over Q_{a}^{2} } P_{a}(z_{a}) 
  T_{0,a}{ {\hat s} \over {\hat s}_{0,a} } , 
\end{equation}
where the sum is taken over all particles in the initial and final 
states (external legs) that can emit the considered radiation.
We call the radiating external leg $a$ the {\it emitter}.
The parameter $\alpha_{s}$ is the QCD coupling, 
and $Q_{a}^{2}$ is the squared momentum transfer of the considered 
branch for the initial-state radiation (ISR) and the squared invariant mass 
of the branch products for the final-state radiation (FSR).
The factor $P_{a}(z_{a})$ represents the corresponding Altarelli-Parisi 
splitting function~\cite{Altarelli:1977zs}.
Although the exact definition depends on the kinematics model~\cite{Odaka:2011hc}, 
the variable $z_{a}$ can be approximately considered as the energy fraction 
carried by the child of the branch, {\it i.e.}, 
the branch product other than the radiation.
Hence, approximately the radiation carries the energy fraction of $1-z_{a}$.
The factor $T_{0,a}$ represents the squared ME of the non-radiative event 
that is reconstructed by removing the considered radiation from the emitter 
{\it a}, by exactly reversing the kinematics model of PS branches.
Finally, the approximation is multiplied by the ratio between the squared 
center-of-mass (CM) energies of the radiative and non-radiative events, 
${\hat s}/{\hat s}_{0,a}$~\cite{Odaka:2012ry}, 
in order to correct for the change of the flux factor in the cross section 
calculation; usually, $\hat{s}/\hat{s}_{0,a} = 1/z_{a}$ for ISR.

\begin{figure}[tp]
\begin{center}
\includegraphics[scale=0.6]{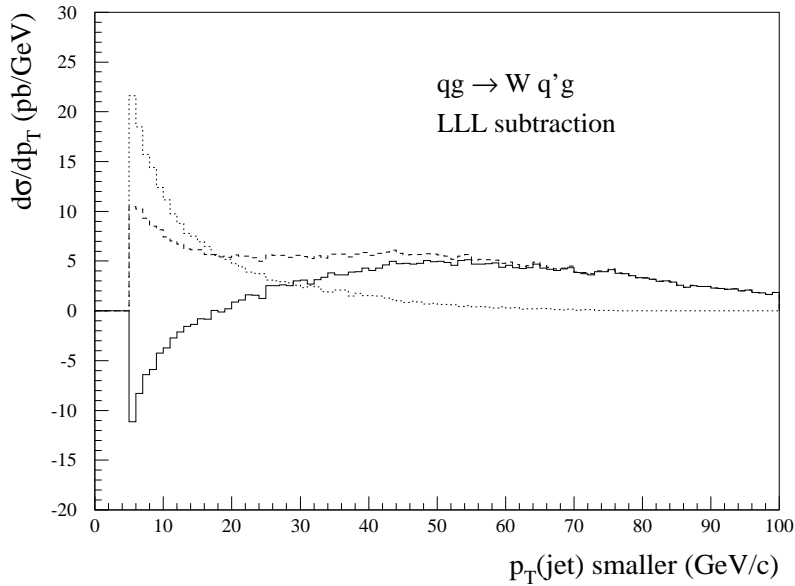}
\caption{\label{fig:llsub}
$p_{T}$ distribution of the smaller $p_{T}$ parton 
in the $qg \rightarrow W q'g$ reaction after the LLL subtraction. 
The simulation is carried out for the 14-TeV LHC condition.
The cut, $p_{T} \geq 5$ GeV and $\Delta R_{jj} \geq 0.2$, is applied to 
cutoff the remaining divergence.
The distribution of positive weight ($+1$) events is illustrated with 
the dashed histogram, while that of negative weight ($-1$) events 
with the dotted histogram.
The unweighted result obtained from the difference between the two distributions 
is shown with the solid histogram.
}
\end{center}
\end{figure}

We have demonstrated in previous studies that the cross sections can be made 
finite by subtracting the approximation in Eq.~(\ref{eq:collapp}) 
from 1-jet MEs~\cite{Odaka:2007gu,Odaka:2011hc}.
However, the subtraction is not successful for 2-jet processes　
based on 2-jet production MEs.
Figure~\ref{fig:llsub} shows the $p_{T}$ distribution of the smaller 
$p_{T}$ parton in the $qg \rightarrow W q'g$ reaction after the subtraction, 
where $p_{T}$ is measured with respect to the beam direction. 
The quark combination, $qq'$, represents all possible pairs of up-type and 
down-type quarks up to the $b$ quark which couple to the $W$ boson 
including CKM non-diagonal combinations.
The simulation has been carried out for the design LHC condition, 
proton-proton collisions at the CM energy of 14 TeV, 
using a leading-order parton distribution function (PDF) 
CTEQ6L1~\cite{Pumplin:2002vw}.
The renormalization scale ($\mu_{R}$) and the factorization scale ($\mu_{F}$) 
are set to the $W$-boson mass ($m_{W}$ = 80.419 GeV).
The cut, $p_{T} \geq 5$ GeV and $\Delta R_{jj} \geq 0.2$, is applied to 
the two final-state partons in order to cutoff the divergence, 
where $\Delta R_{jj}$ is the separation between the two partons 
defined as the quadratic sum of the separations in 
the pseudorapidity and azimuthal angle, 
$\Delta R^{2} = \Delta \eta^{2} + \Delta \phi^{2}$.

In order to deal with 2-jet production MEs, 
we have to subtract the final-state divergence that appears at the limit 
where the two final-state partons are produced collinearly, 
together with the initial-state collinear divergence 
that has been subtracted from 1-jet production MEs. 
We import the subtraction technique that we have developed and 
tested for QED photon radiation in the final state~\cite{Odaka:2012ry}.
The definition of the approximation in Eq.~(\ref{eq:collapp}) is valid 
also for the final-state divergence;
the difference lies only in the definition of the splitting function.

We evaluate the approximation in Eq.~(\ref{eq:collapp}) if the parton 
that is taken as the radiation is judged to be {\it soft}.
The radiation is defined as soft when the $Q$ value for the assumed 
radiation-emitter combination is smaller than a given energy scale.
We define this energy scale to be equal to $m_{W}$ for both ISR and FSR.
The subtraction component is determined by adding the approximations for 
all possible assignments of the radiation.

In general, the non-radiative event that is defined by removing the soft 
radiation may again include a soft radiation when we treat multi-jet events.
In order to make such {\it doubly soft} contributions finite, 
we need to subtract next-to-leading-logarithmic (NLL) components 
together with LL ones.
The evaluation of NLL terms is beyond the scope of the present study.
Hence, we simply ignore their contributions. 
We terminate the evaluation and set the ME value to zero 
once such a {\it doubly soft} combination is found. 

Since the subtraction is unphysical, 
the remaining cross section may become negative in some phase-space regions.
GR@PPA generates events having weights of $+1$ and $-1$ according to 
the sign of the cross section.
The spectra of the positive- and negative-weight events are separately 
shown with dashed and dotted histograms, respectively, in Fig.~\ref{fig:llsub}.
The final spectrum shown with the solid histogram is obtained as 
the difference between them.
The obtained spectrum indicates a negative divergence at $p_{T} \rightarrow 0$.
Namely, the collinear approximation in Eq.~(\ref{eq:collapp}) overestimates 
the divergence in the soft region.

The properties of another divergence, the soft-gluon divergence, 
have been discussed by Catani and Seymour~\cite{Catani:1996vz}.
They have shown that the asymptotic form of this divergence, 
the soft-gluon approximation, can be expressed in a form similar to 
the collinear approximation in Eq.~(\ref{eq:collapp}) as\footnote{
This formula is identical to the $\epsilon \rightarrow 0$ limit of 
Eq.~(4.7) in Ref.~\cite{Catani:1996vz}.}
\begin{equation}\label{eq:softapp}
  T^{\rm soft} = \sum_{a} T^{\rm soft}_{a} , \ \ \
  T^{\rm soft}_{a} = -{ 8\pi\alpha_{s} \over p_{a}q } 
  \left\{ \sum_{b} { ({\bf T}_{a} \cdot {\bf T}_{b})(p_{a}p_{b}) 
  \over (p_{a}+p_{b})q } \right\} T_{0,a} , 
\end{equation}
where $q$ denotes the four-momentum of the considered soft gluon.
In this approximation, we have to consider the {\it spectator} $b$ 
together with the {\it emitter} $a$ because this divergence originates 
from the interference between the radiations from different external legs.
The variables $p_{a}$ and $p_{b}$ represent their four-momenta, 
and ${\bf T}_{a} \cdot {\bf T}_{b}$ is the corresponding color charge.
This divergence emerges in gluon radiations, 
while no such divergence appears in quark radiations.
Starting from this expression, 
Catani and Seymour introduced the dipole subtraction method 
for next-to-leading order calculations, 
in which the proposed subtraction function converges to 
the collinear approximation at the collinear limit 
and to the soft-gluon approximation at the soft limit of the gluon radiation.
In the following, we introduce another form of the subtraction function 
that is suitable for application in our event generator.

Here, we have to note that the color factor is assumed to factorize 
in the expression in Eq.~(\ref{eq:softapp}). 
However, the factorization is not trivial.
The amplitude of a diagram $m$ for jet-production processes can be 
subdivided into a color part $M_{C}^{m}$ and a kinetic part $M_{p}^{m}$. 
If we do not assume the color factorization, 
the soft-gluon approximation can be written as
\begin{equation}\label{eq:softapp2}
  T^{\rm soft}_{a} = -{ 8\pi\alpha_{s} \over p_{a}q } 
  \sum_{b} { p_{a}p_{b} \over (p_{a}+p_{b})q } 
  {\rm Re} \left[ \sum_{m_{0}{\rm ,}n_{0}} 
  \left( M_{C}^{m_{0} \otimes R_{a}} M_{C}^{n_{0} \otimes R_{b} \dagger} \right)
  \left( M_{0{\rm ,}p}^{m_{0}} M_{0{\rm ,}p}^{n_{0} \dagger} \right) \right] ,
\end{equation}
where $M_{C}^{m_{0} \otimes R_{a}}$ represents the color part for the 
radiative diagram that is transformed to the non-radiative diagram $m_{0}$ 
by the removal of a gluon attached to the external leg $a$, 
and the kinetic part for the non-radiative diagram $m_{0}$ is written 
as $M_{0{\rm ,}p}^{m_{0}}$.
If the color part factorizes, {\it i.e.}, the ratio of the color parts, 
\begin{equation}\label{eq:colratio}
  Y_{ab}^{m_{0}n_{0}} = 
  { M_{C}^{m_{0} \otimes R_{a}} M_{C}^{n_{0} \otimes R_{b} \dagger}
  \over M_{0{\rm ,}C}^{m_{0}} M_{0{\rm ,}C}^{n_{0} \dagger } } ,
\end{equation}
is independent of the diagram combination $(m_{0}, n_{0})$, 
the ratio can be considered as the color charge, 
$Y_{ab}^{m_{0}n_{0}} = {\bf T}_{a} \cdot {\bf T}_{b}$,
and the expression in Eq.~(\ref{eq:softapp2}) can be reduced to 
Eq.~(\ref{eq:softapp}).

In general, the color part factorizes only when the number of colored 
external legs included in the non-radiative event is 
two or three~\cite{Catani:1996vz}, 
whereas it does not factorize for more complicated events. 
Actually, we confirmed that the ratio in Eq.~(\ref{eq:colratio}) 
depends on the diagram combination in $W$ + 3 jet production processes 
by explicitly calculating the ratio.
However, we confine ourselves to 2-jet production processes 
in this article.
The color part factorizes in this case.
Hence, in the following, we assume that the soft-gluon divergence can be 
approximated by the function in Eq.~(\ref{eq:softapp}).

First of all, we compare the collinear-limit behavior of the soft-gluon 
approximation with the soft-limit behavior of the collinear 
approximation.
We assume hereafter that all partons are massless.
Hence, the $a = b$ term vanishes in Eq.~(\ref{eq:softapp}).
As the energy fraction of the radiation can be taken as $1 - z_{a}$, 
the four-momentum of the radiation can be approximated as 
$q \rightarrow (1-z_{a})p_{a}$ for ISR and 
$q \rightarrow (1-z_{a})p_{a}/z_{a}$ for FSR at the collinear limit.
Since $Q_{a}^{2} = -(p_{a}-q)^{2}$ for ISR and 
$Q_{a}^{2} = (p_{a}+q)^{2}$ for FSR, 
the denominator, $p_{a}q$, in Eq.~(\ref{eq:softapp}) is equal to 
$Q_{a}^{2}/2$ both in ISR and FSR. 
Therefore, since 
\begin{equation}\label{eq:colsum}
  {\bf T}_{a} \cdot \sum_{b} {\bf T}_{b} = 0 , 
\end{equation}
the soft-gluon approximation in Eq.~(\ref{eq:softapp}) can be further 
approximated as
\begin{equation}\label{eq:soft2coll}
  T^{\rm soft}_{a {\rm (ISR)}} \rightarrow 
  8\pi\alpha_{s} {2{\bf T}_{a}^{2} \over (1-z_{a})Q_{a}^{2} }T_{0,a} , \ \ \  
  T^{\rm soft}_{a {\rm (FSR)}} \rightarrow 
  8\pi\alpha_{s} {2{\bf T}_{a}^{2}z_{a} \over (1-z_{a})Q_{a}^{2} }T_{0,a} , 
\end{equation}
at the collinear limit.

In order to examine the soft-limit behavior of the collinear approximation, 
we define the function ${\hat P}_{a}(z)$ as
\begin{equation}\label{eq:phat}
  P_{a}(z) = { 2{\bf T}_{a}^{2} \over 1-z } {\hat P}_{a {\rm (ISR)}}(z) 
  = { 2{\bf T}_{a}^{2}z \over 1-z } {\hat P}_{a {\rm (FSR)}}(z) . 
\end{equation}
The color factors for gluon radiations are 
${\bf T}_{q \rightarrow qg}^{2} = C_{F} = 4/3$ 
and ${\bf T}_{g \rightarrow gg}^{2} = C_{A} = 3$.
The definitions for ISR and FSR are slightly different from each other.
The explicit definitions are given as 
\begin{equation}\label{eq:phat2}
  {\hat P}_{q \rightarrow qg {\rm (ISR)}}(z) = { 1+z^{2} \over 2 } , \ \ \ 
  {\hat P}_{g \rightarrow gg {\rm (ISR)}}(z) = { \{1-z(1-z)\}^{2} \over z } , 
\end{equation}
and ${\hat P}_{a {\rm (FSR)}}(z) = {\hat P}_{a {\rm (ISR)}}(z)/z$.
Note that all the ${\hat P}_{a}(z)$ functions approach unity 
at the soft limit, $z \rightarrow 1$.
Since ${\hat s}/{\hat s}_{0,a} \rightarrow 1$ at the soft limit, 
we obtain the soft-limit behavior of the collinear approximation in 
Eq.~(\ref{eq:collapp}) as
\begin{equation}\label{eq:coll2soft}
  T^{\rm coll}_{a {\rm (ISR)}} \rightarrow 
  8\pi\alpha_{s} {2{\bf T}_{a}^{2} \over (1-z_{a})Q_{a}^{2} }T_{0,a} , \ \ \  
  T^{\rm coll}_{a {\rm (FSR)}} \rightarrow 
  8\pi\alpha_{s} {2{\bf T}_{a}^{2}z_{a} \over (1-z_{a})Q_{a}^{2} }T_{0,a} . 
\end{equation}
Namely, the collinear limit of the soft-gluon approximation 
is identical to the soft limit of the collinear approximation.

Although the above conclusion may sound trivial, 
we gain a crucial insight into divergence approximations from this discussion.
Let us consider a modified soft-gluon approximation defined as 
\begin{equation}\label{eq:modsoft}
  \tilde{T}^{\rm soft} = \sum_{a} \tilde{T}^{\rm soft}_{a} , \ \ \
  \tilde{T}^{\rm soft}_{a} = T^{\rm soft}_{a}{\hat P}_{a}(z_{a})
  { {\hat s} \over {\hat s}_{0,a} } .
\end{equation}
Since ${\hat P}_{a}(z_{a}) \rightarrow 1$ and 
${\hat s}/{\hat s}_{0,a} \rightarrow 1$ at $z_{a} \rightarrow 1$, 
this approximation approaches the soft-gluon approximation 
at the soft limit.
Simultaneously, because of the property in Eq.~(\ref{eq:soft2coll}),
the approximation in Eq.~(\ref{eq:modsoft}) approaches the collinear 
approximation in Eq.~(\ref{eq:collapp}) at the collinear limit.
Therefore, we must be able to subtract all divergences in multi-jet MEs 
by using the approximation in Eq.~(\ref{eq:modsoft}) 
instead of the collinear approximation in Eq.~(\ref{eq:collapp}).

However, the direct use of the approximation in Eq.~(\ref{eq:modsoft}) 
is problematic; the cross section integration is hard to converge.
We apply the subtraction not only in regions close to the soft limit, 
but also significantly away from the limit.
The approximation in Eq.~(\ref{eq:softapp}) exhibits unexpected behavior 
in the latter.
It is better to eliminate the contribution of the soft-gluon approximation 
away from the soft limit.
As a solution, we introduce a combined approximation 
that is defined as
\begin{equation}\label{eq:combapp}
  T^{\rm comb} = \sum_{a} T^{\rm comb}_{a} , \ \ \
  T^{\rm comb}_{a} = ( 1 - r_{\rm soft} )T^{\rm coll}_{a}
  + r_{\rm soft} \tilde{T}^{\rm soft}_{a} ,
\end{equation}
where $r_{\rm soft}$ is an arbitrary dumping coefficient that should 
approach unity at the soft limit and reduce to zero far away from the limit.
It is obvious that this approximation has the same properties as 
that in Eq.~(\ref{eq:modsoft}) at the soft limit and the collinear limit.
We have tested several definitions of $r_{\rm soft}$ and found that 
the simplest definition, 
\begin{equation}\label{eq:dumpco}
  r_{\rm soft} = z_{a} ,
\end{equation}
provides good convergence properties in all cases that we have tested.
Hence, we use this definition through the following discussions.

It should be noted here that 
if there are only two colored particles in the non-radiative event, 
the soft-gluon approximation in Eq.~(\ref{eq:softapp}) coincides 
with the soft limit of the collinear approximation, 
Eq.~(\ref{eq:coll2soft}), without taking the collinear limit 
if it is evaluated in a frame in which the two colored particles are 
aligned back-to-back.
Thus, it is not necessary to account for the soft-gluon divergence 
separately.
This condition applies to the 1-jet production processes in hadron collisions 
that we have studied previously. 
Other cases where this condition holds include 
the $e^{+}e^{-} \rightarrow q{\bar q}g$ process in the frame in which 
the $q{\bar q}$ pair is produced back-to-back, 
and the deep inelastic scattering, $\ell q \rightarrow \ell qg$, 
in the Breit frame.

\begin{figure}[tp]
\begin{center}
\includegraphics[scale=0.6]{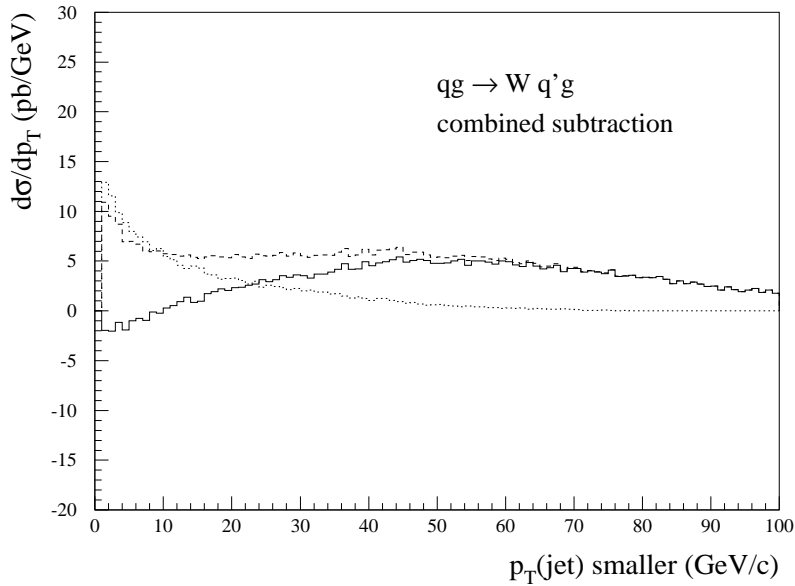}
\caption{\label{fig:combsub}
$p_{T}$ distribution of the smaller $p_{T}$ parton 
in the $qg \rightarrow W q'g$ reaction after the combined subtraction. 
The simulation condition and the histogram notation are same as those 
of Fig.~\ref{fig:llsub}, 
except that the cut for the produced partons is relaxed to 
$p_{T} \geq 1$ GeV and $\Delta R_{jj} \geq 0.01$.
}
\end{center}
\end{figure}

Figure~\ref{fig:combsub} shows the $p_{T}$ spectrum of the smaller $p_{T}$ parton 
in the $qg \rightarrow W q'g$ reaction after the combined subtraction, 
in which the combined approximation defined in Eqs.~(\ref{eq:combapp}) 
and (\ref{eq:dumpco}) is used for the subtraction when the final-state 
gluon is considered as the radiation, while the ordinary LLL subtraction 
is applied when $q'$ is assumed to be the radiation.
The gluon is always treated as the radiation when the FSR component is 
evaluated.
For the calculation of the soft-gluon approximation, 
the color charge can be assigned as 
${\bf T}_{q} \cdot {\bf T}_{q'} = -(C_{F}-C_{A}/2)$ and 
${\bf T}_{q} \cdot {\bf T}_{g} = {\bf T}_{q'} \cdot {\bf T}_{g} = -C_{A}/2$
from the sum rule in Eq.~(\ref{eq:colsum}). 
The total subtraction factor is evaluated by adding all these approximations.
The cut is relaxed to the usual one, 
$p_{T} \geq 1$ GeV and $\Delta R_{jj} \geq 0.01$, 
that is applied only for numerical stability.
The other conditions are same as those for the simulation shown in Fig.~\ref{fig:llsub}.
We can see that the negative divergence at $p_{T} \rightarrow 0$ has 
disappeared although a finite negative offset remains.
This offset might be caused by the mismatch in the kinematics model 
because we are using different models for ISR and FSR.
In any case, even if it is due to the mismatch, 
the existence of such a small offset is harmless 
since the spectrum at small $p_{T}$ is overwhelmed by PS radiations 
from non-radiative events.

\begin{figure}[tp]
\begin{center}
\includegraphics[scale=0.6]{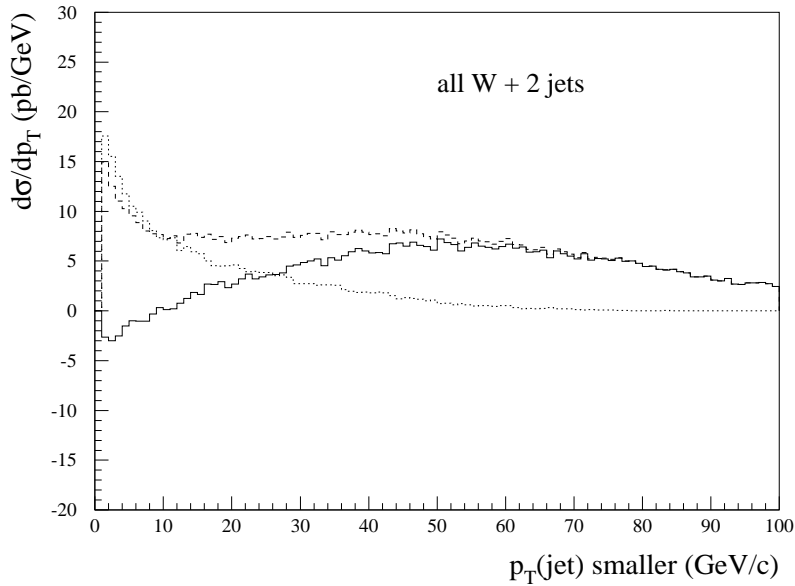}
\caption{\label{fig:allw2j}
$p_{T}$ distribution of the smaller $p_{T}$ parton after the subtraction. 
All $W$ + 2 jet production processes are included.
The combined subtraction is applied to the $qg \rightarrow W q'g$ and 
$q{\bar q}' \rightarrow W gg$ reactions, 
while the LLL subtraction is applied to the other processes 
in which no gluon emerges in the final state.
The simulation condition and the histogram notation are same as those 
of Fig.~\ref{fig:combsub}.
}
\end{center}
\end{figure}

A similar result has also been obtained for another $W$ + 2 jet production 
process, $q{\bar q}' \rightarrow W gg$.
The LLL subtraction provides finite cross sections for the other $W$ + 2 jet 
processes that include no gluon in the final state: 
$gg \rightarrow W q{\bar q}'$, $q{\bar q}' \rightarrow W q''{\bar q}''$, 
$qq' \rightarrow W q''q'$, {\it etc}.
The $p_{T}$ distribution of the smaller $p_{T}$ parton for the combined 
$W$ + 2 jet production process is shown in Fig.~\ref{fig:allw2j}.
The spectrum is similar to that of the $qg \rightarrow W q'g$ subprocess 
in Fig.~\ref{fig:combsub} 
because this subprocess dominates the combined result in the LHC condition.

Since the color factorization is assumed, 
the above discussions are restricted to 2-jet production processes. 
However, the factorization is assumed only for proving the identity 
between the collinear limit of the soft approximation
and the soft limit of the collinear approximation. 
If the identity is generally proved, the above subtraction method can also 
be applied to larger jet-multiplicity processes.
Although it is yet to be formally proved, 
numerical studies indicate that the identity holds in 3-jet production processes.

\section{Soft-gluon correction to PS-applied events}
\label{sec:pscorr}

The subtracted components are restored by PS radiations from lower 
jet-multiplicity events in our matching method.
Since the subtraction is altered by accounting for the soft-gluon divergence, 
the PS simulation should also be altered accordingly.
As we can see in Eq.~(\ref{eq:softapp}),  
the soft-gluon approximation does not fully factorize; 
{\it i.e.}, the kinetic factor depends on the momenta of particles 
composing the non-radiative event.
Therefore, it is impossible to implement the corresponding correction 
as an alteration to PS branches.
Instead, we adopt a method to alter the production frequency 
of lower jet-multiplicity events to which the PS simulation is applied.

The divergent components subtracted from $W$ + 2 jet MEs 
have to be restored by PS radiations from $W$ + 1 jet events.
Hence, we discuss the correction to PS-applied $W$ + 1 jet 
($W$ + 1 jet $\otimes~{\rm PS}$) events in the following.
In this PS simulation, the PS has to be applied to the final-state parton 
as well as the initial-state partons. 
Since the kinematics model of the PS branches plays a key role 
in the ME-PS matching~\cite{Odaka:2007gu}, 
the performance of the initial-state (spacelike) PS model has been 
intensively examined in previous studies on leading-jet 
matching~\cite{Odaka:2009qf,Odaka:2012iz,Odaka:2013fb}.
On the other hand, the performance of the final-state (timelike) PS 
has not been explicitly examined in these studies.
The kinematics model of the final-state PS that is implemented in GR@PPA 
is described in a previous article~\cite{Odaka:2011hc}.
The performance of this model has been studied, 
and necessary corrections to achieve precise matching have been established 
in the study of diphoton production~\cite{Odaka:2012ry} 
by using the QED photon radiation from quarks as the probe. 
We adopt this model and the established corrections in the present study.
A similar timelike PS is also applied to the partons radiated 
in the initial-state PS~\cite{Odaka:2011hc}.

The soft-gluon correction to $W$ + 1 jet $\otimes~{\rm PS}$ events is 
evaluated as follows. 
First of all, we determine the hardest PS branch in each event.
The search is performed over both initial-state and final-state branches.
The hardness is defined by the $p_{T}$ value of each branch, 
where $p_{T}$ is defined as $p_{T}^{2} = (1-z)Q^{2}$ for ISR 
and $p_{T}^{2} = z(1-z)Q^{2}$ for FSR.
Subsequently, we boost the included hard-interaction event to its CM frame, 
and rotate the hardest PS branch so that it aligns along 
the emitter parton in the hard-interaction event 
in order to remove the effects of softer branches.
The rotation is performed so as to minimize the angular change in order to 
preserve the angular information of the radiation as strictly as possible.
From the boosted hard-interaction event and the hardest branch, 
we construct a radiative event following the kinematics model that 
we adopt in PS branches.
We evaluate the collinear and combined approximations for this radiative event, 
and take the ratio, $T^{\rm comb}/T^{\rm coll}$, as the event weight.
The squared ME value is multiplied by this event weight before it is 
passed to the MC integration and event generation utility 
BASES/SPRING~\cite{Kawabata:1985yt,Kawabata:1995th} included in GR@PPA 
using the LabCut framework~\cite{Odaka:2011hc,Odaka:2012ry}. 
The PS simulation is allowed before passing the ME value to BASES/SPRING 
in this framework.
In this way, the evaluated event weight is automatically accounted for 
in the cross-section integration and event generation. 

There are some remarks on the hardest-branch search.
Starting from the hard interaction, 
the search is performed in decreasing order of $Q^{2}$ for both ISR and FSR.
In ISR, the search is done for radiations from spacelike partons only.
In FSR, when PS branches attached to a quark in the hard interaction 
are investigated, 
only those radiations from this quark line are examined, 
while all $g \rightarrow gg$ branches are examined 
when PS branches attached to a gluon are investigated.
The search is terminated if we encounter a flavor-singlet branch, 
$g \rightarrow q\bar{q}$, or reach the end of the branch chain.
We ignore the other branches in order to ignore irreducible higher-order 
effects included in the PS simulation.
For example, 
when PS branches are attached to a $q\bar{q}' \rightarrow W g$ event, 
one of the gluon radiations from the quark that is produced by a 
$g \rightarrow q''\bar{q}''$ branch of the final-state gluon 
may become hardest.
Since the property of the gluon radiation of a quark is different 
from that of a gluon, 
the observed hardest gluon radiation cannot be considered as a radiation 
from the $q\bar{q}' \rightarrow W g$ event; 
instead, it should be attributed to a radiation from 
a $q\bar{q}' \rightarrow W q'' \bar{q}''$ event.
We ignore such higher-order effects by terminating the search 
at the $g \rightarrow q''\bar{q}''$ branch. 
If the flavor-singlet branch is hardest, 
we return the unit event weight because such an event corresponds to 
a quark-radiation event for which we do not need to consider 
the soft divergence.

\begin{figure}[tp]
\begin{center}
\includegraphics[scale=0.5]{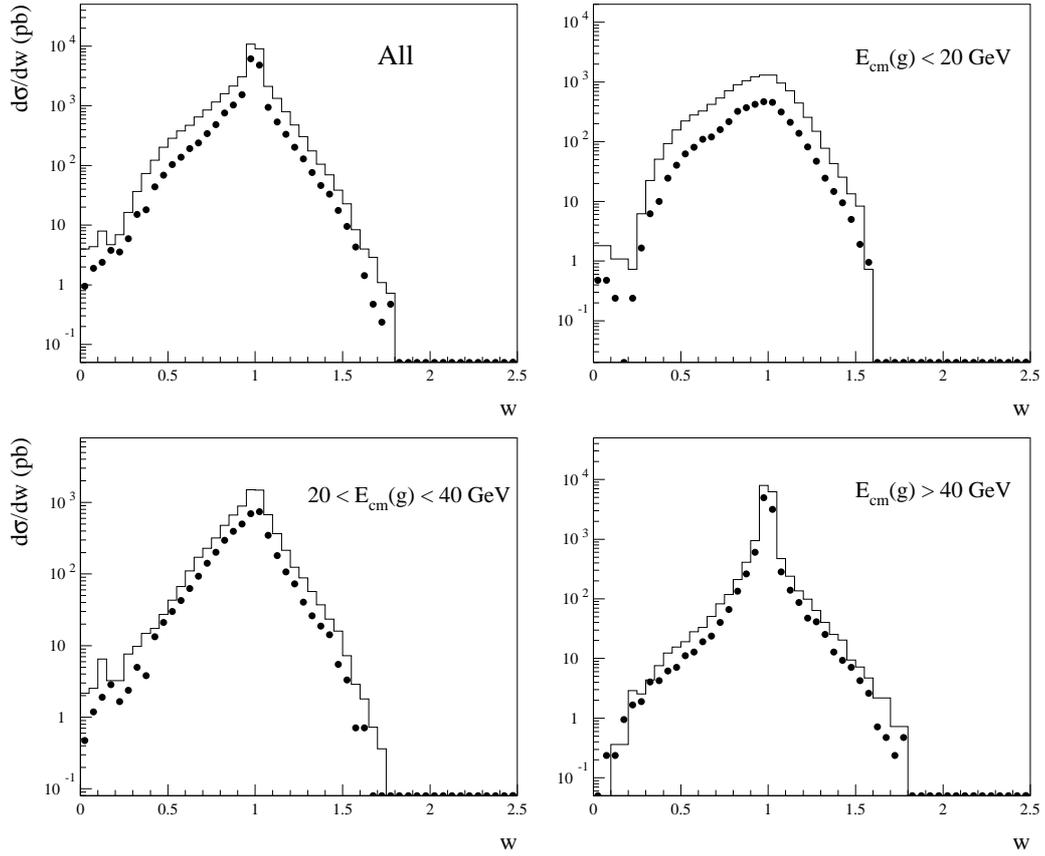}
\caption{\label{fig:sgcorr}
Distribution of the soft-gluon correction factor.
The plots show the distributions of the event weight evaluated for 
$qg \rightarrow W q' \otimes {\rm PS}$ events in which a gluon is selected 
as the hardest radiation in the PS.
The histograms show the $T^{\rm comb}/T^{\rm coll}$ ratio 
of $qg \rightarrow W q'g$ events generated according to the LLL approximation.
The events have been generated in the 14-TeV LHC condition.
The distributions for three $E_{\rm cm}(g)$ regions are presented 
together with that for all events. 
}
\end{center}
\end{figure}

In Fig.~\ref{fig:sgcorr}, the distribution of the event weight evaluated 
for $qg \rightarrow W q' \otimes {\rm PS}$ events, 
in which a gluon is selected as the hardest radiation in PS, 
is compared with the distribution of the $T^{\rm comb}/T^{\rm coll}$ ratio 
of $qg \rightarrow W q'g$ events generated according to the LLL 
approximation.
The event generation has been carried out in the 14-TeV LHC condition.
Although the event weight of $qg \rightarrow W q' \otimes {\rm PS}$ events 
is evaluated, the ME values are not corrected for it in the event generation.
The compared distribution of $qg \rightarrow W q'g$ events shows 
the alteration in the subtraction. 
Therefore, the comparison shows us how precisely the alteration is compensated 
by the correction to $qg \rightarrow W q' \otimes {\rm PS}$ events.

The cuts are carefully arranged to ensure that the two distributions are 
comparable.
The $Q^{2}$ value is required to be larger than $m_{W}^{2}$ 
for $q'$ with respect to the initial-state gluon in both 
$qg \rightarrow W q'$ events in $qg \rightarrow W q' \otimes {\rm PS}$ 
and non-radiative $qg \rightarrow W q'$ events reconstructed 
in the LLL approximation to $qg \rightarrow W q'g$.
A $p_{T}$ condition, $p_{T} \geq 5$ GeV, is required to the hardest branch 
in $qg \rightarrow W q'\otimes{\rm PS}$. 
Accordingly, the same condition is required to the gluon in the 
$qg \rightarrow W q'g$ events with respect to the initial-state partons 
and to the combined momentum of the two final-state partons.

The two distributions compared in Fig.~\ref{fig:sgcorr} are in good agreement 
in their shapes 
although there is a significant difference in the normalization.
The difference in the normalization is due to the difference 
in the gluon energy spectrum.
The spectrum of the hardest PS branch is suppressed at low energy 
as an effect of multiple radiation.
The comparison is also made in Fig.~\ref{fig:sgcorr} in three regions 
of the gluon energy in the CM frame, $E_{\rm cm}(g)$.
We can see that the difference in the normalization is large at low energy 
and becomes smaller at high energy, 
while the agreement in the shape holds in all regions.
Hence, we expect that the thus evaluated correction 
to the $qg \rightarrow W q' \otimes {\rm PS}$ events 
will reasonably compensate for the alteration in the subtraction 
from $qg \rightarrow W q'g$ MEs in visible high-energy regions.
The possible difference in the normalization, 
particularly in low-energy regions, 
can be attributed to a higher-order effect which we do not need to worry about.
In any case, the soft-gluon correction is at most 10\% on average 
even for very soft gluon radiations.
Therefore, even though the soft-gluon correction plays a significant role 
in the subtraction of divergent components from $W$ + 2 jet MEs, 
its impact to the finite $W$ + 1 jet cross section is limited.

\section{Matched event generation}
\label{sec:matchedgen}

The performance of the matching method described in previous sections 
is tested for $W$-boson production in the LHC condition in the following.
We combine $W$ + 0, 1, and 2 jet production events generated 
according to $W$ + 0, 1, and 2 jet MEs, respectively, 
using the GR@PPA event generator.
The $W$ bosons are assumed to decay to a pair of electron and neutrino
and the PS simulation is fully applied down to $Q = 5.0$ GeV.
The combined subtraction described in Sec.~\ref{sec:combsub} is applied 
in the generation of $W$ + 2 jet events, while the LLL subtraction and 
the soft-gluon correction described in Sec.~\ref{sec:pscorr} 
are applied in the generation of $W$ + 1 jet events.
The non-divergent contribution of {\it doubly soft} events is ignored 
in the generation of $W$ + 2 jet events as described in Sec.~\ref{sec:combsub}.
The soft-gluon correction is applied 
not only to $qg \rightarrow W q' \otimes {\rm PS}$ events 
but also to $q{\bar q}' \rightarrow W g \otimes {\rm PS}$ events 
in order to compensate for the alteration in the subtraction 
from $q{\bar q}' \rightarrow W gg$ MEs.
On the other hand, no correction is applied to $W$ + 0 jet events.
The other generation conditions are same as those of previous simulations 
in this article, except that the collision CM energy is decreased to 7 TeV 
in order to compare the result with experimental data~\cite{Aad:2012en}.

As described in Sec.~\ref{sec:combsub}, the energy scales, 
the renormalization scale ($\mu_{R}$), factorization scale ($\mu_{F}$), 
and scales for ISR ($\mu_{\rm ISR}$) and FSR ($\mu_{\rm FSR}$), 
are all set to $m_{W}$ as the default.
The scales $\mu_{\rm ISR}$ and $\mu_{\rm FSR}$ define the maximum $Q^{2}$ of 
the PS and subtraction in the initial and final states, respectively.
Although this choice may not be appropriate for high-$p_{T}$ jet production, 
fixed energy scales are convenient for testing the matching properties 
since the relation to visible quantities such as the jet $p_{T}$ is 
straightforward.

The generated events are processed by PYTHIA 6.425~\cite{Sjostrand:2006za} 
to add softer interactions and hadronization simulations.
The default setting in PYTHIA is unchanged except for the settings of 
{\tt PARP(67) = 1.0} and {\tt PARP(71) = 1.0}, 
as in the previous studies~\cite{Odaka:2009qf,Odaka:2012iz,Odaka:2013fb}.
The event selection and the jet reconstruction are applied to 
the obtained hadron-level events according to the definition in 
the ATLAS measurement~\cite{Aad:2012en}.
The events are accepted if the electron and neutrino from the $W$ decay 
satisfy the following conditions on the transverse momentum ($p_{T}$), 
pseudorapidity ($\eta$), and transverse mass ($m_{T}$),
\begin{equation}\label{eq:evsel}
  p_{T}^{e} > 20 {\rm \ GeV} , \ \ \ 
  |\eta^{e}| < 2.5, \ \ \ 
  p_{T}^{\nu} > 25 {\rm \ GeV} , \ \ \ 
  {\rm and} \ \ \ m_{T} > 40 {\rm \ GeV} ,
\end{equation}
where $m_{T}$ is defined as 
$m_{T}^{2} = 2(p_{T}^{e}p_{T}^{\nu} - {\vec p}_{T}^{e}{\vec p}_{T}^{\nu})$.
The {\it Born-level} momentum of the electron before applying 
the photon-radiation correction is used for this selection.
The hadron jets are reconstructed 
by using FastJet 3.0.3~\cite{Cacciari:2011ma}, 
with the application of the anti-$k_{T}$ algorithm with $R = 0.4$.
All stable particles, except for the electron from the $W$ decay 
and neutrinos, are used for the reconstruction.
The reconstructed jets satisfying the condition, 
\begin{equation}\label{eq:jetsel}
  p_{T}^{{\rm jet}} > 30 {\rm \ GeV} , \ \ \ 
  |\eta^{{\rm jet}}| < 4.4 , \ \ \ 
  {\rm and} \ \ \ \Delta R(e,{\rm jet}) > 0.5 ,
\end{equation}
are assumed to be detected, where $\Delta R(e,{\rm jet})$ denotes 
the separation from the decay electron in $\Delta R$.

\begin{figure}[tp]
\begin{center}
\includegraphics[scale=0.6]{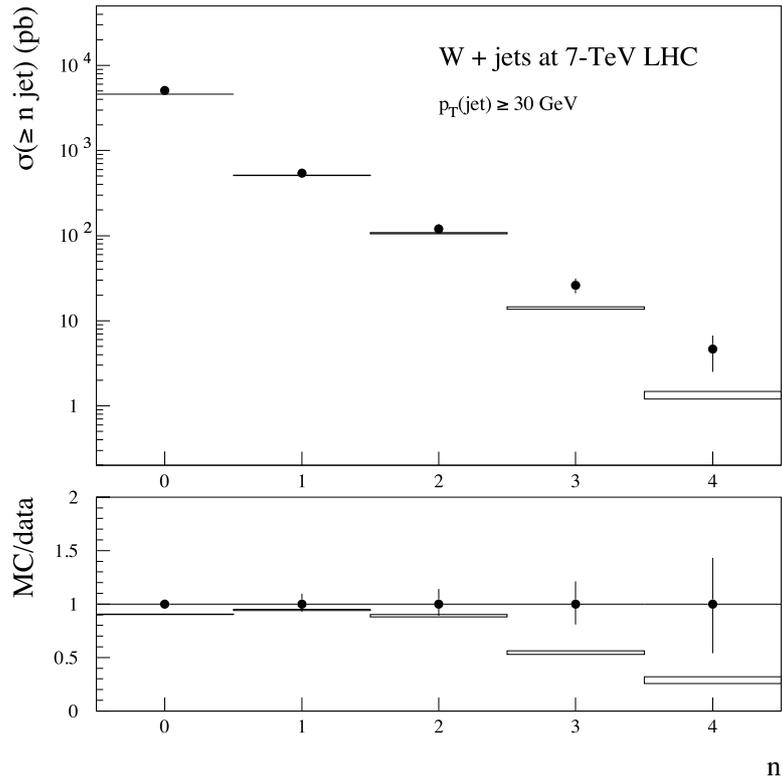}
\caption{\label{fig:njinc}
Integrated cross section as a function of the threshold jet multiplicity.
The ATLAS measurement at LHC is plotted with error bars, 
while the simulation result described in the text is shown with boxes.
The vertical size of the boxes indicates the statistical error of 
the simulation. 
}
\end{center}
\end{figure}

The cross section obtained from the simulation is presented as a function 
of the threshold jet multiplicity in Fig.~\ref{fig:njinc}. 
The statistical error of the simulation is illustrated using boxes. 
The result is compared with the corresponding ATLAS measurement 
at 7 TeV~\cite{Aad:2012en} plotted with error bars. 
It is to be noted that no error bar is drawn for the measurement 
in the first bin because the measurement value is not available in the data 
archive\footnote{{\tt http://durpdg.dur.ac.uk/view/ins1083318/short}.}.
The plotted value, which corresponds to the total cross section, 
has been derived from the $\sigma(\geq 1 {\rm ~jet})$ result and 
the result for the $\sigma(\geq 1 {\rm ~jet})/\sigma(\geq 0 {\rm ~jet})$ ratio 
available in the archive.

We can see that the simulation is in good agreement with the measurement 
up to the threshold jet multiplicity of two 
and significantly underestimates the measurement for higher multiplicities.
This result is reasonable since we have fully evaluated the jet production 
up to two jets in the simulation.
Events with three or more jets are generated by PS radiation 
covering a limited $Q^{2}$ range.
It should be noted that, 
since the MEs for the event generation are all evaluated at the tree level, 
there is no guarantee concerning the absolute value of the cross section.
The data/simulation ratio in the total cross section that is shown 
in the first bin is the so-called K-factor.
The agreement up to two jets is further improved 
if we correct the normalization of the simulation with this factor (= 1.11).

\begin{figure}[tp]
\begin{center}
\includegraphics[scale=0.6]{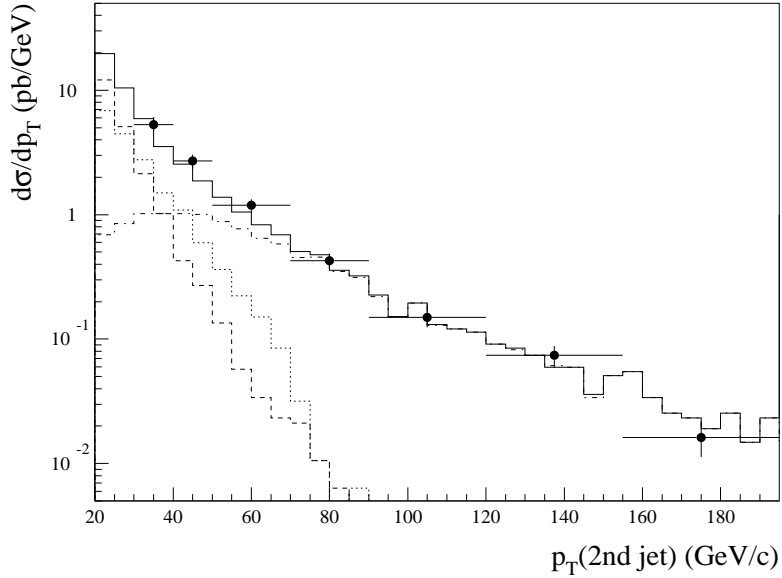}
\caption{\label{fig:ptj2}
$p_{T}$ distribution of the second jet.
The simulation results are illustrated with histograms to be compared 
with the ATLAS measurement plotted with error bars.
Together with the total yield (solid), the contributions from $w0j$ (dashed), 
$w1j$ (dotted) and $w2j$ (dot-dashed) are separately shown.
The jet-$p_{T}$ threshold of the simulation is lowered to 20 GeV/$c$ 
in order to show the low-$p_{T}$ behavior.
}
\end{center}
\end{figure}

The matching in the combination of $W$ + 0 and 1 jet events has already 
been discussed in previous studies.
The soft-gluon correction applied to $W$ + 1 jet events does not have 
significant impact on the discussion.
The matching of the newly included $W$ + 2 jet process can be studied 
by investigating the property of the second jet,
where the detected jets are ordered according to the $p_{T}$ value. 
Figure~\ref{fig:ptj2} shows the $p_{T}$ spectrum of the second jet. 
Together with the summed spectrum, 
the contributions from $W$ + 0 jet ($w0j$), 1 jet ($w1j$) and 2 jet ($w2j$) 
events are separately presented.
We can see that the contributions from $w0j$ and $w1j$, 
which are predominantly determined by PS, are dominant at low $p_{T}$, 
while $w2j$ is dominant at high $p_{T}$. 
The long high-$p_{T}$ tail of $w2j$ shows a characteristic power-law 
behavior expected from $W$ + 2 jet MEs.
The sum of the three contributions shows a smooth transition 
from the low- to high-$p_{T}$ regions, to yield a spectrum 
which is in good agreement with the plotted ATLAS measurement.

\begin{figure}[tp]
\begin{center}
\includegraphics[scale=0.5]{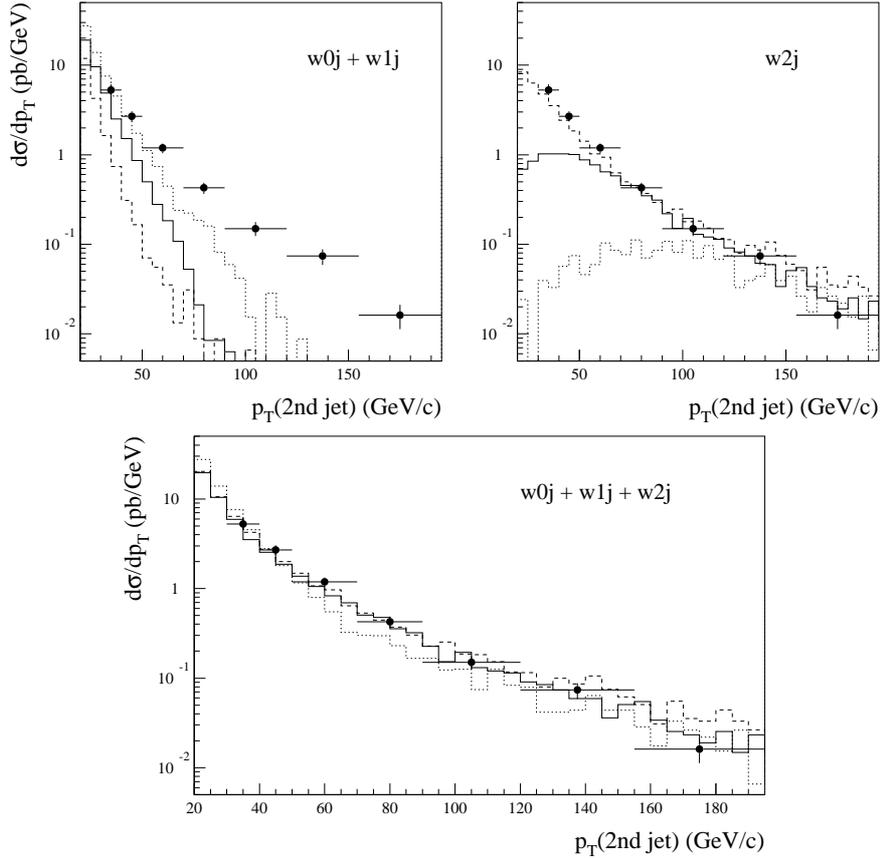}
\caption{\label{fig:ptj2-sdep}
$\mu_{F}$ dependence of the second-jet $p_{T}$ distribution.
The simulation results are shown with histograms: 
the results for $\mu_{F} = m_{W}/2$ (dashed) and $2m_{W}$ (dotted) 
together with the default result with $\mu_{F} = m_{W}$ (solid).
The ATLAS measurement is plotted to guide the comparison.
The $w0j + w1j$ and $w2j$ contributions are separately shown in upper panels, 
and the combined results are shown in the lower panel.
}
\end{center}
\end{figure}

Although the smooth jet-$p_{T}$ spectrum observed in the above is already 
satisfactory evidence for the ME-PS matching, 
a further test can be carried out by investigating the stability of 
the spectrum against variation in the energy scales.
The second-jet $p_{T}$ spectra for the choice of $\mu_{F} = m_{W}/2$ and 
$\mu_{F} = 2m_{W}$ are presented in Fig.~\ref{fig:ptj2-sdep}, 
together with the default result with $\mu_{F} = m_{W}$.
The renormalization scale $\mu_{R}$ for ME calculations is fixed 
to $m_{W}$ in this study. 
The other scales, $\mu_{\rm ISR}$ and $\mu_{\rm FSR}$, are set equal to 
$\mu_{F}$ to test the matching.
We can see apparent $\mu_{F}$ dependences in the separate $w0j + w1j$ 
and $w2j$ contributions. 
These dependences compensate each other in the combined result 
to yield a smooth spectrum which is significantly less dependent on $\mu_{F}$.

\begin{figure}[tp]
\begin{center}
\includegraphics[scale=0.5]{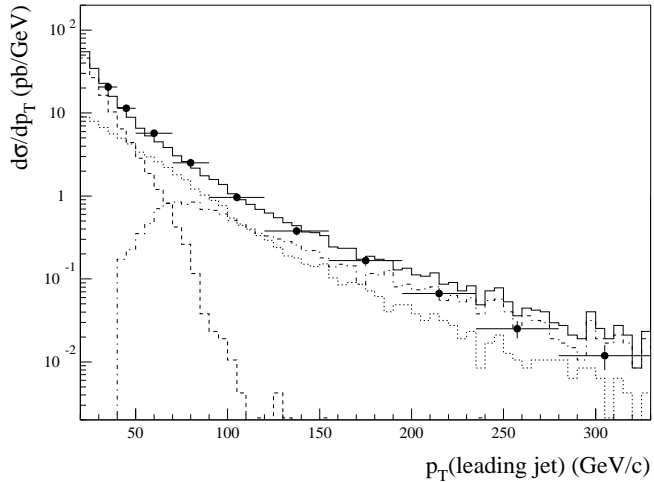}
\caption{\label{fig:ptj1}
$p_{T}$ distribution of the leading jet.
The simulation results are illustrated with histograms to be compared 
with the ATLAS measurement plotted with error bars.
Together with the total yield (solid), the contributions from $w0j$ (dashed), 
$w1j$ (dotted) and $w2j$ (dot-dashed) are separately shown.
The jet-$p_{T}$ threshold of the simulation is again lowered to 20 GeV/$c$. 
}
\end{center}
\end{figure}

The $w2j$ contribution has an apparent scale dependence since 
the inclusion of the divergent LL component varies according to $\mu_{F}$.
This dependence may induce a scale dependence in the $p_{T}$ spectra 
of the leading jet and $W$.
Figure~\ref{fig:ptj1} shows the $p_{T}$ spectrum of the leading jet, 
{\it i.e.}, the highest-$p_{T}$ jet.
The contributions from $w0j$, $w1j$ and $w2j$ are separately shown 
together with the summed spectrum.
We naively expected that the leading-jet $p_{T}$ would be 
predominantly determined by the $w1j$ contribution at high $p_{T}$.
However, as we can see in Fig.~\ref{fig:ptj1},
the $w2j$ contribution is comparable to or larger than the $w1j$ contribution 
at $p_{T} \gtrsim \mu_{F}$.
Hence, the induced scale dependence may become significant 
in this $p_{T}$ spectrum.

\begin{figure}[tp]
\begin{center}
\includegraphics[scale=0.5]{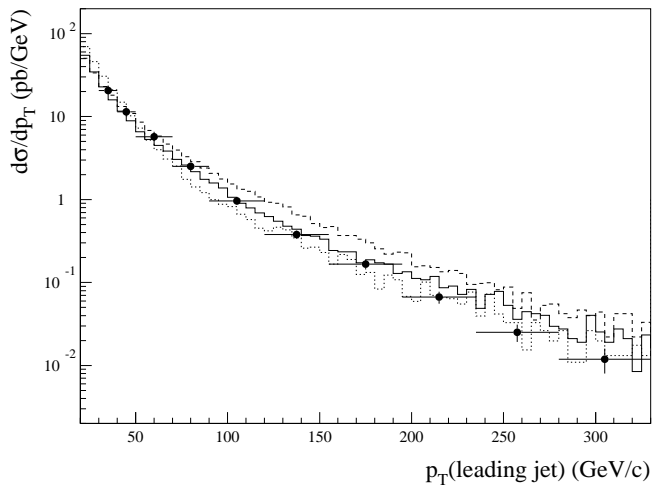}
\caption{\label{fig:ptj1-sdep}
$\mu_{F}$ dependence of the leading-jet $p_{T}$ distribution.
The simulation results are shown with histograms
and compared with the ATLAS measurement plotted with error bars. 
The simulation results for $\mu_{F} = m_{W}/2$ (dashed), 
$m_{W}$ (solid), and $2m_{W}$ (dotted) are presented.
}
\end{center}
\end{figure}

The $\mu_{F}$ dependence of the leading-jet $p_{T}$ spectrum is shown 
in Fig.~\ref{fig:ptj1-sdep}.
Although the simulation shows reasonable agreement with the measurement,
the $\mu_{F}$ dependence is not small at high $p_{T}$ as we were concerned.
A similar amount of $\mu_{F}$ dependence is also observed in the 
$W$-boson $p_{T}$ spectrum since the $W$-boson $p_{T}$ is predominantly 
determined by the leading jet at high $p_{T}$.
In order to obtain a stable spectrum, 
the unavoidable $\mu_{F}$ dependence of $w2j$ has to be compensated 
with the $\mu_{F}$ dependence of the $w1j$ contribution,
which can be merely induced by the $\mu_{F}$ dependence of PDF.
However, the $w1j$ contribution is very stable against $\mu_{F}$ variation 
in the relevant $p_{T}$ range.
This stability is accidental since the $\mu_{F}$ dependence of PDF 
is determined by the combination of the QCD evolution and the functional 
form of PDF, {\it i.e.}, the dependence on the momentum fraction of partons. 
The $\mu_{F}$ dependence observed in Fig.~\ref{fig:ptj1-sdep} is thus 
unavoidable in our matching method.

As we can see in Figs.~\ref{fig:ptj2-sdep} and \ref{fig:ptj1-sdep}, 
the simulation tends to yield larger cross sections than the measurement 
at high $p_{T}$.
This tendency may be attributed to the inappropriate choice of 
the energy scales, particularly to the small $\mu_{R}$ value (= $m_{W}$) 
compared to the jet $p_{T}$.
The capability of the simulation as a measurement tool will be discussed 
elsewhere by carrying out a detailed comparison with measurements.
A more appropriate definition of the energy scales will be pursued 
in the course of the study.

\section{Conclusion}
\label{sec:concl}

We have extended the LLL subtraction method for ME-PS matching, 
which we have developed for combining 0-jet and 1-jet production processes 
in association with electroweak-boson production in hadron collisions, 
to 2-jet production processes.
We have introduced an approximation that simultaneously reproduces 
the collinear divergence and the soft-gluon divergence 
in 2-jet production MEs.
This approximation is used for the subtraction instead of the 
collinear approximation in the LLL subtraction 
in order to make the 2-jet MEs finite.
The alteration in the subtraction can be compensated by applying 
an appropriate correction to 1-jet events 
to which the PS simulation is applied to generate additional jets.
This matching method can be generalized for processes including three 
or more jets if the identity between the collinear limit of the soft 
approximation and the soft limit of the collinear approximation 
is proved.

The performance of the developed matching method has been tested 
using $W$ + 0, 1, and 2 jet production processes, 
and the simulation result has been compared with the ATLAS measurement 
at LHC.
The inclusion of the 2-jet process shows a remarkable impact to 
the second-jet $p_{T}$ spectrum. 
The 2-jet contribution exhibiting a long tail to high $p_{T}$ is smoothly 
connected to low-$p_{T}$ contributions from 0-jet and 1-jet processes, 
yielding a spectrum in good agreement with the measurement.

The contribution from the 2-jet process has an apparent factorization-scale 
($\mu_{F}$) dependence in our matching method.
We have shown that this dependence is compensated by the $\mu_{F}$ dependence 
of the 0-jet and 1-jet contributions in the second-jet $p_{T}$ spectrum.
On the other hand, 
this $\mu_{F}$ dependence induces a sizable $\mu_{F}$ dependence 
in the leading-jet $p_{T}$ spectrum at high $p_{T}$.
This is caused by the fact that the $\mu_{F}$ dependence of the $W$ + 1 jet 
contribution at high $p_{T}$, which is determined by PDF, 
is accidentally very small.

\section*{Acknowledgments}

This work has been carried out as an activity of the NLO Working Group, 
a collaboration between the Japanese ATLAS group and the numerical analysis 
group (Minami-Tateya group) at KEK.
The authors wish to acknowledge useful discussions with the members, 
particularly K. Kato of Kogakuin University.

\providecommand{\href}[2]{#2}\begingroup\raggedright\endgroup

\end{document}